\documentclass[rmp,aps,preprint,showpacs,showkeys,preprintnumbers,amsmath,amssymb,floatfix]{revtex4-1}

\usepackage{graphicx}
\usepackage{dcolumn}
\usepackage{bm}
\usepackage{epsfig}
\usepackage{xcolor}
\usepackage{stmaryrd} 

\usepackage{amsmath, nccmath}
\usepackage{geometry}

\usepackage{amssymb}
\usepackage{mathtools}
\usepackage{braket}
\usepackage{stackengine}
\usepackage[version=3]{mhchem} 

\DeclarePairedDelimiter\floor{\lfloor}{\rfloor}

\begin{document}

\preprint{Gerbi-et-al-2022-Appl. Surf. Sci.-
https://doi.org/10.1016/j.apsusc.2022.155218
}

\title{Phase-Space Ab-Initio Direct and Reverse 
Ballistic-Electron Emission Spectroscopy:
Schottky Barriers Determination for Au/Ge(100)}

\author{Andrea Gerbi}
\email{andrea.gerbi@spin.cnr.it}
\author{Renato Buzio}
\email{renato.buzio@spin.cnr.it}
\affiliation{CNR-SPIN Institute for Superconductors, I-16152 Genova (Italy)}
\author{Cesar Gonz\'{a}lez}
\email{cesar.gonzalez@ucm.es}
\affiliation{Departamento de F\'isica de Materiales (U. Complutense), E-28040 Madrid (Spain)}
\author{Fernando Flores}
\email{fernando.flores@uam.es}
\affiliation{F\'isica Te\'orica de la Materia Condensada-IFIMAC (U. Autonoma), E-28049 Cantoblanco (Spain)}
\author{Pedro L. de Andres$^{*}$}
\email{pedro.deandres@csic.es}
\affiliation{Instituto de Ciencia de Materiales de Madrid (CSIC), E-28049 Cantoblanco (Spain)}
\affiliation{$^{*}$ On leave of absence at nanotech{@}surfaces Laboratory, Swiss Federal Laboratories for Materials Science and Technology, CH- 8600 D{\"u}bendorf (Switzerland)}

\date{\today}

\begin{abstract}
We develop a phase-space ab-initio formalism to compute Ballistic Electron Emission Spectroscopy current-voltage I(V)'s in a metal-semiconductor interface. We consider injection of electrons into the conduction band for direct bias ($V>0$) and injection of holes into the valence band or injection of secondary Auger electrons into the conduction band for reverse bias ($V<0$).
Here, an ab-initio description of the semiconductor inversion layer (spanning hundreds of Angstroms) is needed.
Such formalism is helpful to get parameter-free best-fit values for the Schottky barrier, a key technological characteristic for metal-semiconductor rectifying interfaces. We have applied the theory to characterize the Au/Ge(001) interface; a double barrier is found for electrons injected into the conduction band -- either directly or created by the Auger process -- while only a single barrier has been identified for holes injected into the valence band. 
\end{abstract}


\maketitle

\section{Introduction}

An accurate value for the metal-semiconductor Schottky barrier ($\Phi$) is of paramount importance to characterize the rectifying performance of solid-state semiconductor devices~\cite{sze}. 
Furthermore, the increasing trend in miniaturization highlights the need for techniques that can provide values for $\Phi$ related to the microscopic active Schottky region of interest in each case.
Ballistic Electron Emission Microscopy (BEEM) is the foremost technique to obtain precise microscopic values for the Schottky 
barrier~\cite{bellYkaiser88,Moller2008}, and to study the
influence of interface coupling on the electronic
properties~\cite{Cook2015,Wong2020},
but also to study magnetic materials, organic layers, electronic band structure and hot-carrier scattering~\cite{Bell2016} and molecular semiconductors~\cite{Zhou2021}. 

\begin{figure}[!t]
\centering
\includegraphics[width=0.460\linewidth]{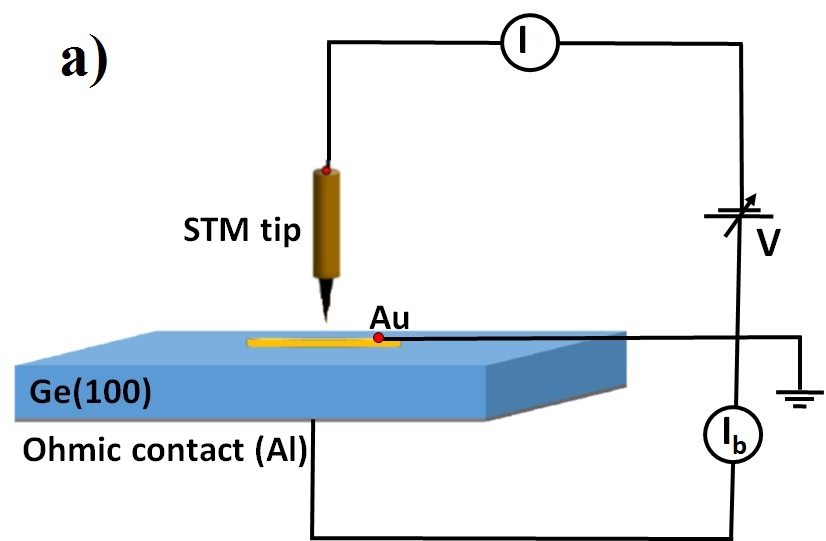}
\includegraphics[width=0.490\linewidth]{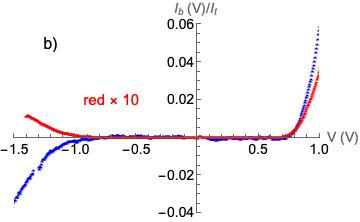}
\includegraphics[width=0.235\linewidth]{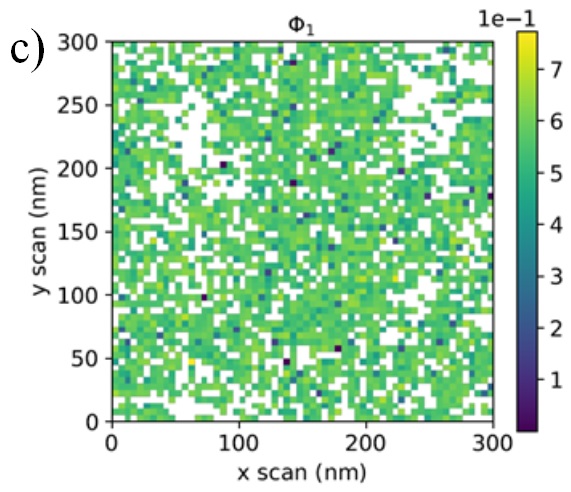}
\includegraphics[width=0.24\linewidth]{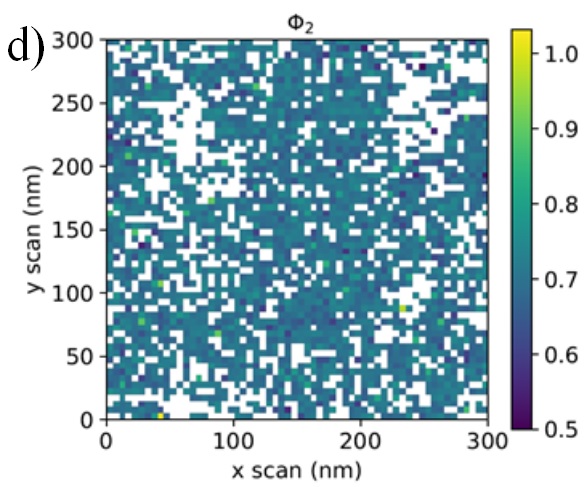}
\includegraphics[width=0.24\linewidth]{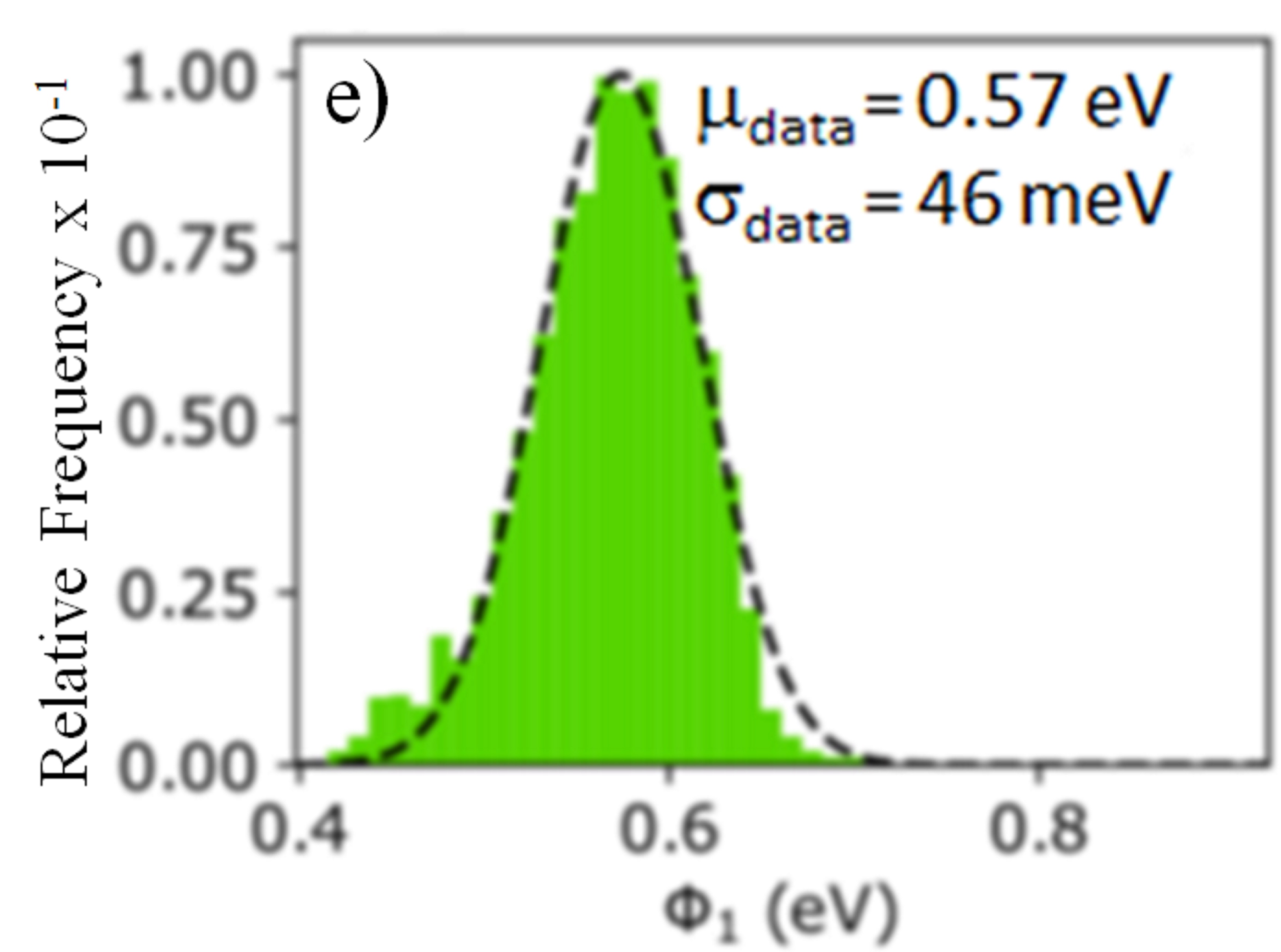}
\includegraphics[width=0.24\linewidth]{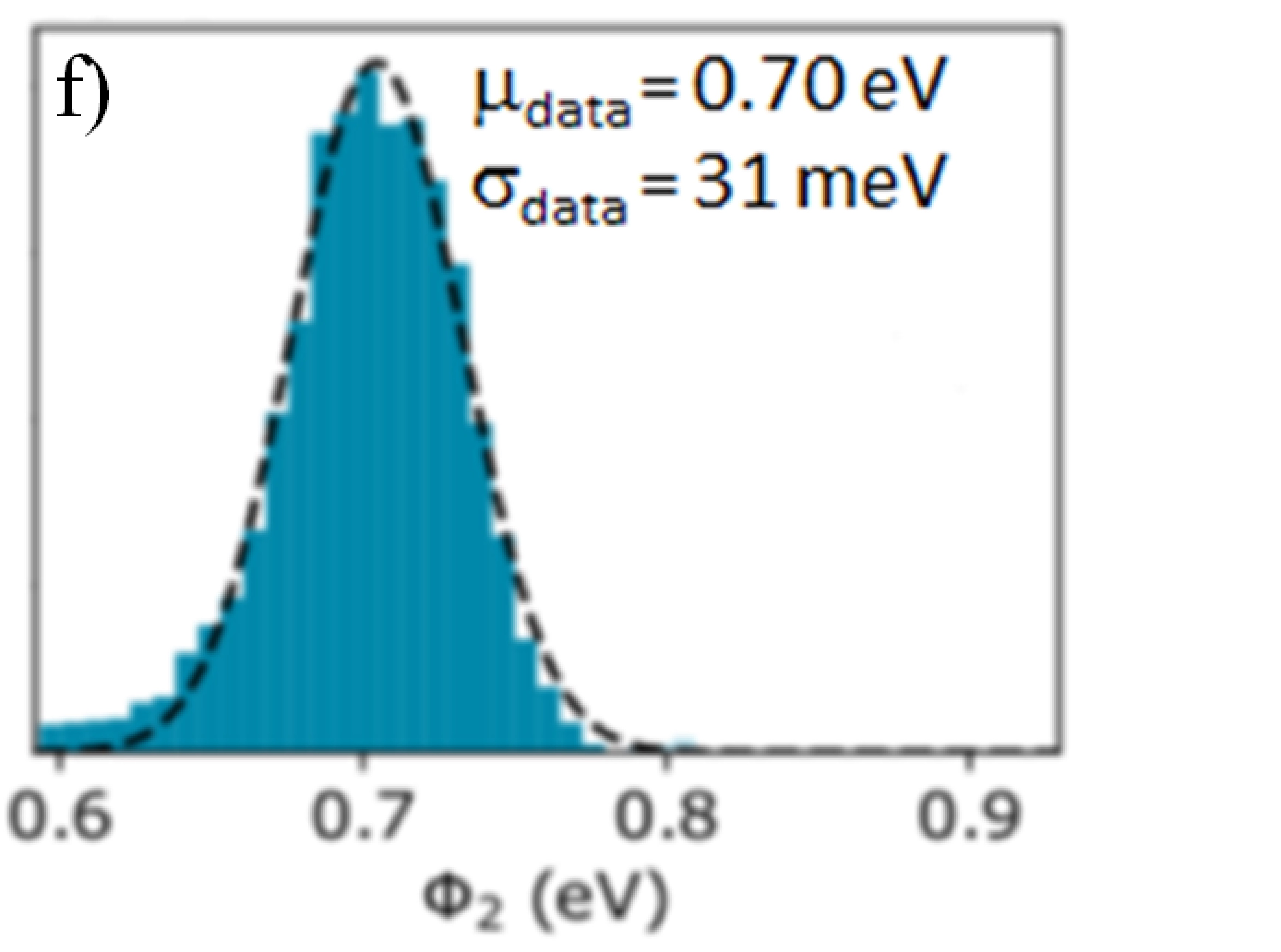}
\caption{
(a) Schematic diagram of the Schottky barrier device with the experimental setup for BEEM measurements. 
(b) The blue curve represents intensity acquired on a region where injection of holes dominates in the reverse bias region RBEEM (blue, spectrum 1, $I_t=2$ nA). It corresponds 
to 34 averaged curves extracted from a grid of 100 curves over an area of $300 \times 300$ nm$^{2}$ and acquired at the edge of the diode for a high doping level sample and a thin Au contact (5-7nm).
The Red curve corresponds to 100 averaged curves acquired on a lower doping level sample with Au thickness in the range of 10-20 nm extracted from a grid of 3600 curves on an area of $300 \times 300$ nm$^{2}$.
Here, Auger-like injection (spectrum 5, $I_t=2$ nA) is enough to nearly cancel the negative current of
holes (notice the factor $10$ used in the plot to show its structure in the same scale as the one
corresponding to holes). 
Both curves were acquired at 80K under dark and UHV condition.
(c,d) Spatially resolved BEEM map of $\Phi_1$ and $\Phi_2$, respectively acquired over a representative Au region in the case of Auger-like injected electrons ($I = 2$ nA, $T= 80$ K). (e,f) Schottky barrier histogram distributions for the two barriers centered at $\Phi_1=0.57$ eV and $\Phi_2=0.70$ eV, respectively. The black lines represent the Gaussian fits to a phenomenological Bell-Kaiser double barrier model with $m=4$. 
}
\label{fgr:fig1}
\end{figure}

BEEM is a scanning tunnelling microscopy (STM)-based technique that can directly measure the barrier height at the metal/semiconductor interface with spatial nanoscale 
resolution~\cite{YI2009,Moller2009,moeller2012,Troadec2012,Parui2013,LaBella2016}. 
The STM tip at bias $V$ injects ballistic electrons into a thin metal overlayer at constant tunnelling current $I_t$ (Figure~\ref{fgr:fig1}-a). When the electron's kinetic energy $e\mid V\mid$ of the injected electrons is higher than the local Schottky barrier formed between the metal and the semiconductor, the electrons can overcome the local barrier and a current $I_b$ is transmitted across the sample and collected through the backside Ohmic contact (Figure~\ref{fgr:fig1}-a)~\cite{Wu2013,Janardhanam2019}.
Accurate Ballistic Electron Emission Spectroscopy (BEES) determination for the Schottky threshold $\Phi$ relies on a best-fit where experimental measurements are compared with a theoretical model where currently the state-of-the-art standard claims accuracies between 10 and 30 meV~\cite{gerbi2020,LaBella2021}.
Such a procedure has been successfully used before in other fields, like structural work related to diffraction techniques, and requires the use of theory as accurate as possible and free of adjustable parameters to avoid spurious correlations in the determination of the value of $\Phi$.
A serious description of BEES must consider four basic steps: (i) tunnelling from the STM tip to the metal base, (ii) transport through the metal base, (iii) transmission and reflection at the interface and (iv) injection into the semiconductor conduction ($V>0$, electrons) or valence ($V<0$, holes) bands. The Keldysh non-equilibrium Greens functions formalism developed in~\cite{claveau2017} performs such tasks. Such theory has been complemented with an ab-initio LCAO procedure~\cite{Lewis2011} to build the relevant Hamiltonians making the calculations free from a particular parametrization and more general~\cite{gerbi2018}. Further, the whole procedure has been used to analyse experiments on Ge/Au and Ge/Pt for the injection of electrons on the semiconductor conduction band ($V>0$, direct BEES)~\cite{gerbi2020}. 

Such analysis is incomplete if the injection of holes in the semiconductor valence band is not included (i.e. reverse BEES, $V<0$)~\cite{Balsano2013,Filatov2014,Filatov2018}. In this paper, we extend our previous work to such a new domain. The analysis of reverse experimental I(V) curves shows the necessity to include two new processes that were neglected for $V>0$: 
\begin{enumerate}
\item
The energy losses by quasi-elastic electron-phonon interaction in the depletion layer and, 
\item 
The formation of electron-hole pairs due to secondary Auger-like inelastic electron-electron interaction.
\end{enumerate}

The value of the effective phase-space volume, $\mu$, includes an {\it ab-initio} calculation for each pair of metal-semiconductor interfaces that has to be determined independently for the conduction and valence band. In addition, we derive the modification of $\mu$ because of the aforementioned
two inelastic processes. $\mu$ is approximately increased by $4$ because of phonons and by $2$ because of Auger electrons. Finally, attenuation due to inelastic electron-electron scattering and multiple reflections in the metal base has been thoroughly analyzed in the literature and are necessary ingredients to a complete theoretical description~\cite{reuter2000,Parui2013}.

In summarizing, the formalism presented here provides a simple expression to compute I(V)'s for
the whole domain of voltages within the phase-space approach. It returns accurate parameter-free values of Schottky barriers for positive and negative bias,
allowing the characterization of Schottky barriers with respect to both the conduction and valence bands. Therefore, we have generalized the previous phenomenological ideas by Kaiser and Bell and Ludeke~\cite{bellJVST91,PhysRevLett.70.214}, providing an improved level of analysis for experimental data. Finally, as a by-product of the work done to check and illustrate our theory, we study germanium as a promising material
for high performance devices~\cite{Saraswat2011,Scappucci2013}
and we obtain relevant values for the technologically important Schottky interface Au/Ge by comparing with experiments.

\section{Experimental}
\label{sct:exp}

BEES IVs have been taken on Au/Ge(100) with Ge substrates at three different doping levels (n-type, Sb-doped, MTI Corporation). 
Samples were cut into pieces of 10 mm $\times$ 5 mm sizes and cleaned with acetone and 2-propanol. In order to remove the native oxide, we immersed the samples in hot water at $85^{\circ}$ C, 
Furthermore, we immediately dried it with nitrogen. We then dipped it in an HF solution at 3\% to remove the residual \ce{GeOx} not soluble in hot water and to obtain a hydrogen-terminated Ge(100) surface. 
The cleaned Ge pieces were loaded within a few minutes into the UHV deposition chamber for the Au contact fabrication, obtained by Physical Vapor Deposition (PVD) from a tungsten coil through a shadow mask (vacuum pressure $<10^{-7}$ torr, the deposition rate was $\approx 1.5$ nm min$^{-1}$). A representative set of five averaged spectra was considered. Spectra 1 and 2 correspond to the high doping regime n-Ge case,
i.e. $0.01$ to $0.001$ $\Omega$-cm. The three other spectra correspond to different n-Ge doping regimes:
(i) $0.01$ to $0.001$ $\Omega$-cm for spectrum 3 ($\approx 10^{18}$ cm$^{-3}$, high-doping regime),  (ii) $0.26$-$0.2$ $\Omega$-cm for spectrum 4 ($\approx 10^{16}$ cm$^{-3}$, medium-doping regime) and (iii), $4.46$-$3.97$ $\Omega$-cm for spectrum 5 ($\approx 5\times 10^{14}$ cm$^{-3}$, low-doping regime). 
The nominal thicknesses of the metal contacts was in the range $10$ to $20$ nm for the low doping level samples and it was intentionally kept smaller, $\approx 5$ to $7$ nm, for the high doping level. 
As discussed below, the doping level and the metal contact thickness are critical parameters that control the detection of ballistic holes or the generation of secondary Auger-like electrons in BEES reverse bias voltage.
The contact area was 2.4 mm$^{2}$ for all devices. The Ohmic back contact was fabricated by depositing a thick Al film by pulsed laser deposition at room temperature from a high-purity target~\cite{gerbi2014,buzio2018}. Immediately after contact with the active area of the diode was established 
we have transferred the sample to the UHM LT-STM chamber for BEEM measurements.
We have performed the experiments using a modified commercial STM equipped with an additional low-noise variable-gain current amplifier~\cite{gerbi2018,buzio2020}. Data were taken in the dark and at $T=80$ K to improve the signal-to-noise ratio. For the acquisition of each BEEM spectrum, the tip voltage $V$ was ramped under feedback control, keeping the tunnelling current constant. Noise current fluctuations in individual raw spectra amounted to
$\approx 12$ fA rms. We find that such a low noise level is required for the accurate determination of Schottky barriers by comparing theory and experiment. 

\section{Results and discussion}

\subsection{BEES analysis from the phenomenological Bell-Kaiser model}

The local Schottky barrier, $\Phi$, is obtained from a best-fit analysis of the variation of the collector current in a small interval near the onset, $I_b$ versus $V$, both for the direct BEEM (DBEEM) and for the reverse BEEM (RBEEM). In Figure~\ref{fgr:fig1}-b, we show representative spectra acquired on two different devices, where the hole injection (blue curve, high doping regime) or Auger-like injection (red curve, low doping regime)  are the dominant mechanisms in reverse polarization ($I_b(V<0)<0$ and $I_b(V<0)>0$, respectively). 
Briefly, we have observed hole injection only for (i) diodes prepared on high doping substrates
and, (ii) at the edges of the active region of the diode, where the metal thickness is reduced just
to a few nm. Under such specific conditions we have observed
the inversion of the current ($I_b<0$) in the RBEEM spectra
for about 25\% of the cases.
However, we were unable to spatially resolve the phenomenon. 
In all other situations, the Auger-like injection mechanism ($I_b>0$) was dominant in RBEEM spectra.
According to previous phenomenological ideas by Kaiser and Bell, and Ludeke~\cite{bellYkaiser88,ludeke93JVST}, in the case of Auger-like injection one can get quantitative
information on the local Schottky barrier height by fitting an ensemble of about $2700$ raw
spectra acquired on a grid of $60\times 60$ points (Figures~\ref{fgr:fig1}-c,d), 
using a double barrier Bell-Kaiser model with n=4 for RBEEM,
$I_b(V) = \alpha_1 (V-\Phi_1)^4 + \alpha_2 (V-\Phi_2)^4$.
We estimate from histogram analysis a double barrier distribution centred
at $0.57$ eV and $0.70$ eV (Figures~\ref{fgr:fig1}-e,f), with spatially resolved statistics. These values are slightly different from the full theoretical calculation we present below because they are obtained by comparing against a phenomenological model which uses a mere approximation to the actual volume in phase-space available for
injection of carriers and, it does not include the effect of temperature. 
Nevertheless, such model provides a quick way to analyze data
and as we shall see, it is remarkably close to a more complete theoretical description.

\subsection{Phase-Space formalism}
\label{sct:thPH}

We continue our phase-space ballistic-electron formalism for the direct bias ($V>0$) injection of electrons from the tip to the semiconductor band~\cite{gerbi2020} to the inverse polarization domain ($V<0$). First, we obtain the effective phase-space volume near the Schottky onset for injecting into the semiconductor electrons (conduction band) or holes (valence band). These values are obtained from an ab-initio Keldysh's Greens functions calculation. They are represented by a characteristic value of $\mu$ in Eq. (1), which only depends on the particular metal-semiconductor interface pair. 

Inelastic processes merely attenuate the intensity of electrons injected into the conduction band ($I>0$), which can be easily accounted for by introducing an exponential factor that depends on a characteristic mean free path for electrons, $e^{-\frac{L}{\lambda}}$. The scenario for reverse bias, however, becomes more involved. First, the current of holes ($I<0$ for $V<0$) can only be detected for highly doped samples ($n \ge 10^{18}$ cm$^{-3}$). Indeed, the current of holes injected into the semiconductor valence band under reverse bias voltage is heavily attenuated w.r.t. the current of ballistic electrons for positive voltage, as 
demonstrated by taking the ratio of experimental intensities
$\frac{I_b(+1)}{\mid I_b(-1) \mid}>>1$ (e.g., cf. Figs~\ref{fgr:M36S1} and~\ref{fgr:M42S2}). As we shall see, quasi-elastic phonon-hole interaction inside the inversion layer is mainly responsible for such behaviour.
Second, the appearance of Auger-like electron-hole pairs excitations create a secondary current of electrons injecting into the conduction band ($I>0$) that eventually dominate the negative current of holes injecting into the valence band (e.g., cf. Figs~\ref{fgr:M08S3}, ~\ref{fgr:M02S4}, and~\ref{fgr:M28S5}). 
As explained in~\ref{sct:muph} and~\ref{sct:tauQP}, we take into account
those effects by modifying the effective
phase-space volume by adding a power-law term $E^{\mu'}$ in Equation~\ref{eqn:3}.
For the interaction of ballistic holes with phonons, we obtain $\mu'=4$,
while for the generation of secondary Auger electrons, we get $\mu'=2$.
Therefore, the phase-space formula for the reverse BEES $I_b(V)$ 
can again be written in a simple closed-form as,

$$
I_b(V)=
I_{<}+
\alpha \int_{-\infty}^{0}
E^{\mu'}
\frac{ (E)^{\mu-1}}{1+e^{-\frac{E-(V-\Phi)}{k_B T}}} ~ d E =
$$
\begin{equation}
= I_{<}+
\alpha  \Gamma\left(\mu+\mu'\right) (k_B T)^{\mu+\mu'} \text{Li}_{\mu+\mu'}
\left( -e^{-\frac{V -\Phi}{k_B T}}\right)~;~ V,~\Phi \le 0
\label{eqn:3}
\end{equation}
\noindent
where, 
$\Phi$ is the value for the onset (Schottky barrier) to be sought (eV),
$V-\Phi$ is the tip voltage measured w.r.t to the onset (eV),
$T$ is the absolute temperature (K), 
$\alpha=I_b(V=-\Phi)$ is a proportionality constant used
to normalise the current, 
$\mu$ is the effective dimension of the equivalent ballistic phase space determined
from ab-initio calculations, 
$\mu'$ determines a power-law approximation to the probability of scattering
with other quasiparticles, e.g. phonons or secondary Auger electrons, as discussed
in detail in~\ref{sct:muh} and~\ref{sct:tauQP},
$I_{<}$ is a baseline current below the onset due to noise fluctuations,
and $\Gamma$ and $\text{Li}$ are the incomplete Euler Gamma function
and the Jonqui{\`e}re's function respectively~\cite{polylog}.

One advantage of subsuming an accurate but complex ab-initio calculation in a simple analytical formula such as Eq.~\ref{eqn:3} is the flexibility to include new physical effects. The importance of interfaces in performing devices cannot be overstated, and their continuing influence is foreseen in new challenging domains like spintronics~\cite{Herve2013a,Herve2013b} and devices based on molecular semiconductors, in particular the determination of the metal/molecule energy and the electronic transport gap~\cite{Zhou2021}. In that direction, the next step to developing Eq.~\ref{eqn:3} will be to study spin-polarized effects. 

\begin{figure}[!t]
\centering
\includegraphics[width=0.30\linewidth]{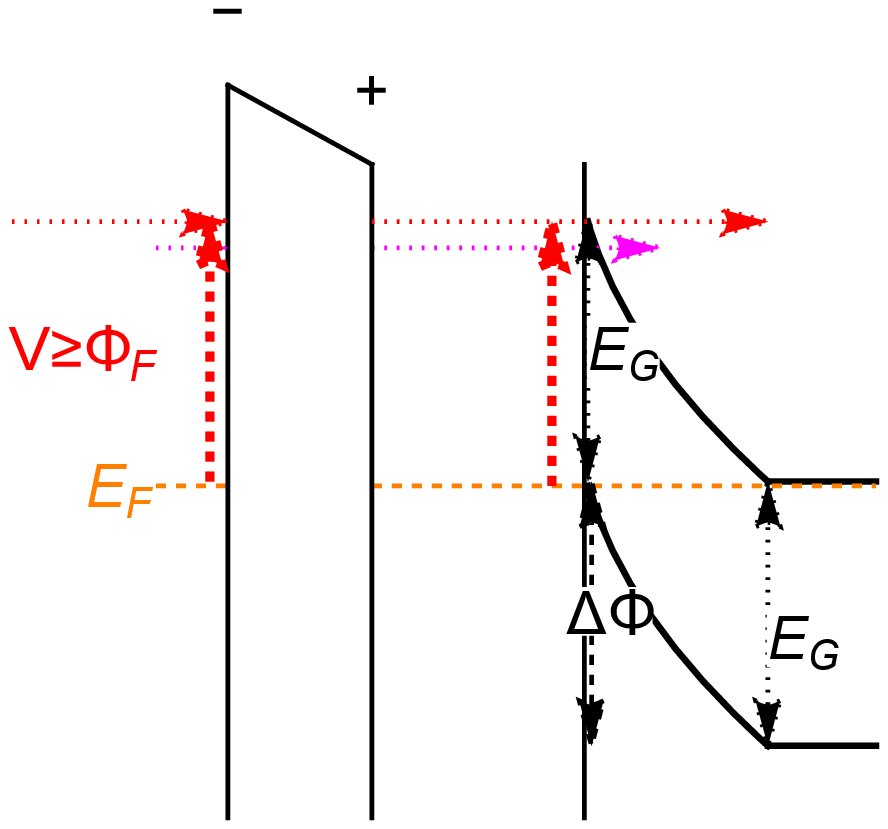}
\includegraphics[width=0.30\linewidth]{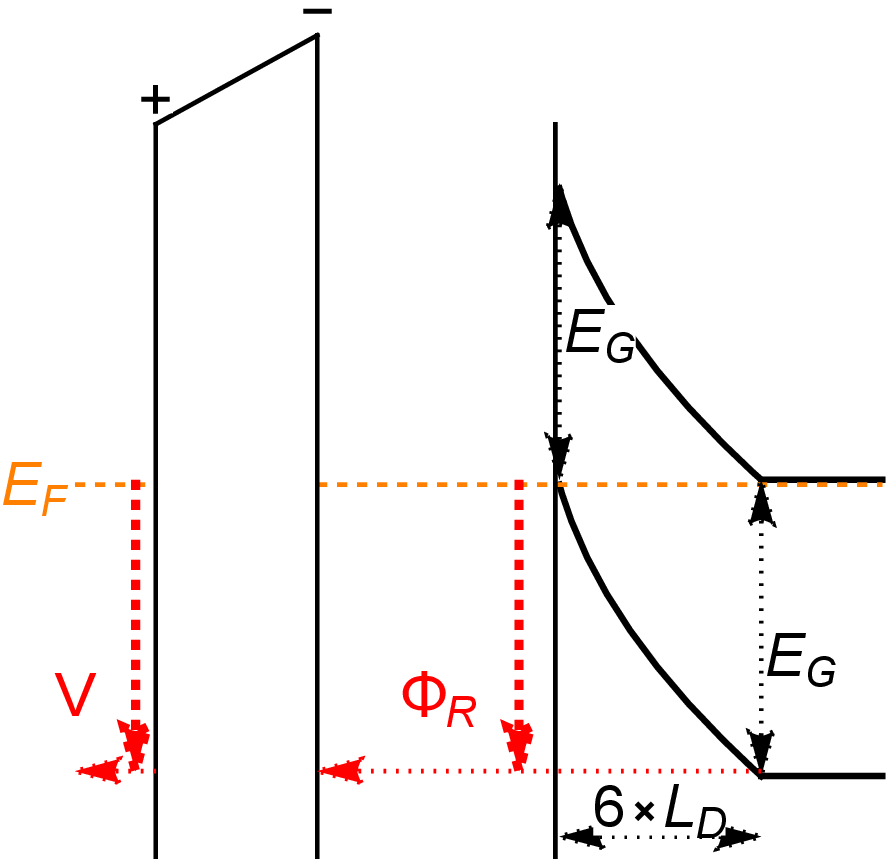} \\
\includegraphics[width=0.30\linewidth]{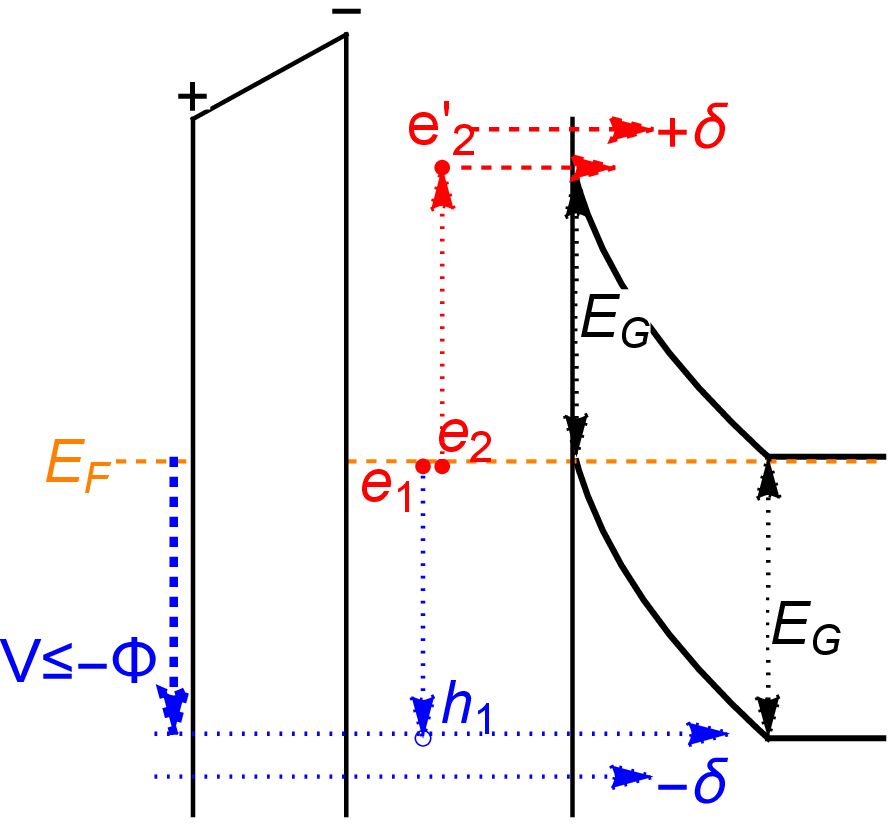}
\includegraphics[width=0.30\linewidth]{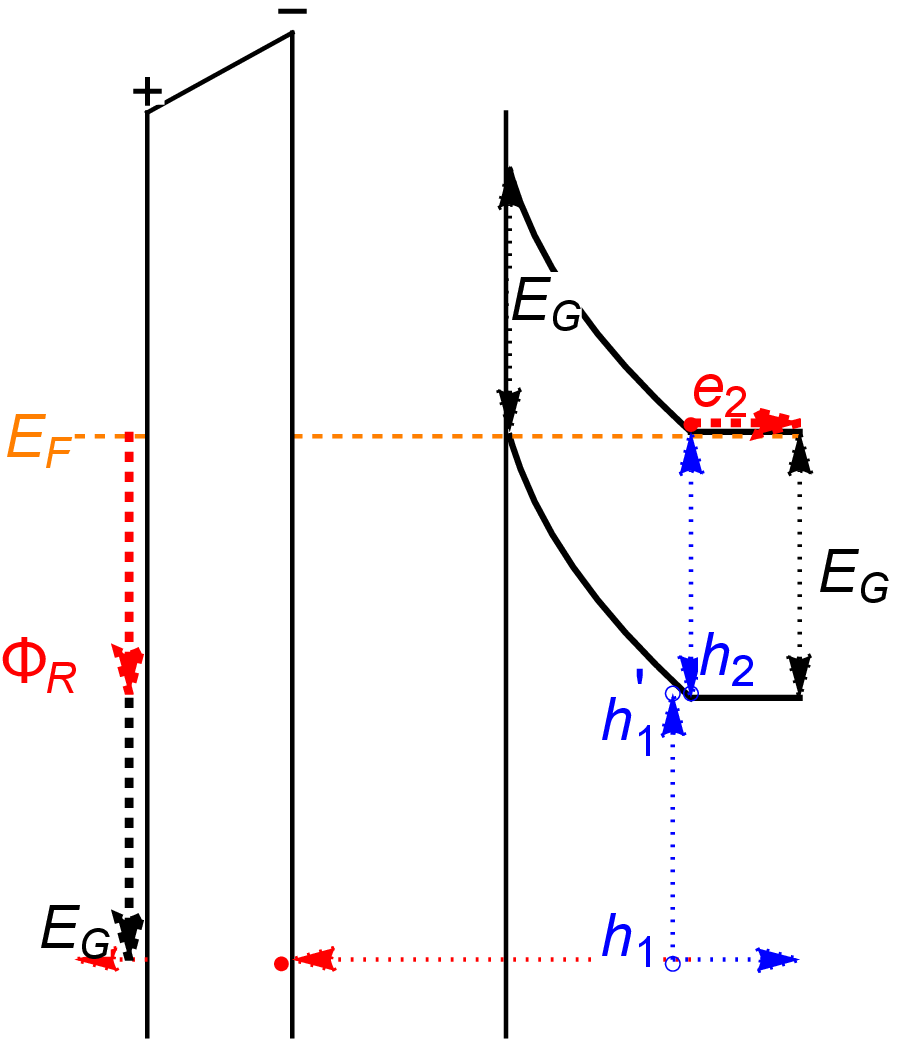}
\caption{
Bands scheme. {\bf Top panels}: Direct and reverse ballistic contributions. 
{\bf Top-Left}. Red: forward ballistic injection of electrons from tip to semiconductor ($I>0$, $V>\Phi_{F}>0$). Magenta: tunneling through the depletion layer (neglected).
{\bf Top-Right}. Red: reverse ballistic injection of electrons from semiconductor to tip ($I<0$ for $V<-\Phi_{R}<0$).
Possible injection from the depletion layer is also neglected.
{\bf Bottom panels}: Reverse Auger processes. 
{\bf Bottom-Left}: 
Auger inelastic injection of electrons from tip to semiconductor 
under reverse bias condiction (($I>0$, $V<-\Phi_{R}$): 
hole $h_1$ (blue) with enough energy to
be injected in the semiconductor valence band decays to $h'_1$
$\Delta E \approx E_G$; the lost energy is used to create the pair $e_2$-$h_2$,
where the electron $e_2$ (red) is injected in the semiconductor's conduction band.
{\bf Bottom-Right}: As in the left panel, but the 
Auger process takes place in the semiconductor, requiring that $V < -(\Phi_{R}+E_{G})$). 
Symbols: $V$ tip bias, $\Phi_{F}$ and $\Phi_{R}$ are forward and reverse Schottky barriers,
$E_G$ semiconductor bandgap, $\Delta \Phi$ is band bending at the surface ($\approx \Phi$ since
the Fermi level lies close to the band edges), orange dashed line 
represents the common Fermi level (chemical potential), the depletion layer is estimated
as approximately $6$ times the Debye length, $L_D$.
}
\label{fgr:bands}
\end{figure}

\subsection{BEES analysis from phase-space formalism}

We have analysed results for five different ensemble-averaged spectra, acquired at specific spots
of the Schottky metal pad.
Spectra 1 and 2 correspond to the high doping case ($0.01$ to $0.001$ $\Omega$-cm) and to the thin regions nearby the edge of the pad, where electron-electron inelastic effects for reverse injection are negligible. Therefore, the injection of holes in the valence band dominates the reverse current. On the other hand, spectra 3, 4 and 5 correspond to the central regions of the metal pad and show strong Auger-like strong re-injection of electrons in the semiconductor conduction band under reverse polarisation. 

\subsubsection{Injection of electrons in the semiconductor conduction band (direct polarisation)}
\label{sct:V0I0}

First, let us consider the injection of electrons from the tip into empty states in the semiconductor conduction band (cf. Figure 2, top-left diagram, $I>0$ for $V>0$). As a reasonable approximation, we neglect tunnelling through the space-charge layer owing to its extension. Thermal excitation of carriers is described by the Fermi-Dirac distribution, which is embedded in our analytical result in Eq. (1). Therefore, for direct bias ($V>0$), we are only concerned with the injection of electrons in the available empty states in the conduction band. In Section~\ref{sct:Vm0I0}, we shall consider the injection of secondary Auger electrons, also into the conduction band, making a positive current for negative bias polarization.

We apply the ab-initio approach developed in~\cite{gerbi2018,gerbi2020};
for Au/Ge we have $\mu=2.11$, resulting in the values in Table~\ref{tbl:VpIp}
(for comparison, in the Pt/Ge we get $\mu=2.01$ and
for the Au/Si interface we have determined $\mu=2.45$).
The main finding is the values for two onsets, at 
$\overline{\Phi_{1}}=0.59\pm 0.03$ and $\overline{\Phi_{2}}=0.70\pm 0.02$ eV. 
The second onset for spectra 3, 4 and 5 draws off about ten times more electrons than the first one, as judged by the ratio between intensities $\frac{\alpha_2}{\alpha_1}$. On the other hand, for the spectra  1 and 2, the ratio goes up to 30 or 60. 
In a previous paper, we have extensively discussed the role of interfacial structural contributions to the origin of the double barrier~\cite{gerbi2020}. Both ab-initio modelling and X-ray diffraction support the existence of two different atomic registries at the metal-semiconductor interface resulting in a shift in the Schottky barriers of $\Delta=0.061$ eV. Such value is in good agreement with our experimental findings. Nonetheless, using our current BEES analysis only, we cannot confidently exclude electronic effects associated with different conduction band minima as suggested in~\cite{prietschYludeke91}.

The attenuation of currents of ballistic electrons is related to the thickness of the metal base layer. It can be estimated by considering the ratio of intensities between a thin and a thick sample. As an example, we take the red and blue curves in Figure~\ref{fgr:fig1}-b
($I_{R}$ and $I_{B}$ respectively) . 
We get for the ratio of experimental intensities $\frac{I_{R}}{I_{B}} = 0.062 \pm 0.002$,
where the mean and the standard deviation have been
estimated by considering the small interval $(0.95,1.05)$ eV around $V=1$ eV. 
This value can be reconciled with the following simple model: a theoretical mean free path in Au,
$\lambda (+1 \text{eV}) \approx 400$ {\AA}, an estimated length difference between
the thick and the thin samples of $L_{R}-L_{B} \approx 150$ {\AA}, a transmission factor 
at the metal-semiconductor interface of $T\approx 0.1$, and five multiple internal reflections\cite{reuter2000},
\begin{equation}
\sum_{i=0,4} (1-i~T)T e^{-\frac{(2i+1)(L_{R}-L_{B})}{\lambda}} \approx 0.06
\end{equation}


\begin{table}[!h]
\begin{tabular}{|c|c|c|c|} \hline
S   &$\Phi_1$&   $\Phi_2$             & $R^2$   \\ \hline       
1   & $0.57\pm 0.04$    & $0.711\pm0.001$    &  0.9995 \\   
2   & $0.59\pm0.01$     & $0.717\pm0.002$     &  0.9986 \\     
3   & $0.58\pm0.01$   &  $0.679\pm 0.001$  &  0.9996 \\ 
4   & $0.60\pm 0.01$    &  $0.686\pm 0.003$   &  0.9994 \\ 
5   & $0.64\pm 0.01$  &  $0.720\pm 0.001$   &  0.9999 \\ 
A   & $0.59\pm 0.03$   & $0.70\pm 0.02$    &          \\\hline
\end{tabular}
\caption{Direct polarization injection of electrons from the tip to the semiconductor conduction band ($V>0$, $I>0$) for spectra S from 1 to 5.
Best fits between experimental values ($T=80$ K) and
Eq.~\ref{eqn:3} for spectra $S=1-5$.  
$\Phi_1 (eV)$ and $\Phi_2 (eV)$ correspond to the first and second thresholds derived from the fits. Standard Errors and $R^2$ for individual cases give a measure of the quality of the fits. The error for the average gives the dispersion.
}
\label{tbl:VpIp}
\end{table}

\subsubsection{Injection of holes in the semiconductor valence band (reverse polarisation)}
\label{sct:Vm0Im0}

Second, we consider the injection of holes into the semiconductor valence band ($V<0$ and $I<0$), demonstrated in spectra 1 and 2. 
As comented while discussing Equation~\ref{eqn:3}, 
the reverse current of holes is weaker than the direct current of electrons. 
The main scattering mechanism responsible for such a strong attenuation is
the quasi-elastic interaction with phonons in the inversion layer.
The attenuation ratio can vary in a wide range depending on the characteristics of the
sample, e.g. from $\frac{I_b(+1)}{\mid I_b(-1) \mid}=40$ in Figure~\ref{fgr:fig1}-b (blue curve) to $400$ in Figures~\ref{fgr:M36S1} and~\ref{fgr:M42S2}. 
Such sensitivity is related to the short mean free path for holes.
We have computed in Section~\ref{sct:lamGe} that for a doping of $10^{18}$ cm$^{3}$
we have $\lambda_{ph} \approx 15$ {\AA}
and a typical width for the inversion layer of $150$ {\AA}.
These values depend significantly on the amount of doping and, 
for a higher doping of $10^{19}$ cm$^{3}$ the inversion layer
becomes merely $50$ {\AA}.
Therefore, the need to include the attenuation due to phonons to describe
ballistic currents of holes, as explained in the discussion of
Equation~\ref{eqn:3}.
The following stringent conditions must be attained to observe the reverse current of holes: high doping, low temperature, and low noise-to-signal ratio.
As discussed in the next section, a short metal base width is also necessary 
to minimize the generation of secondary Auger electrons that oppose and cancel the
current of holes.
For intermediate and low dopings of $10^{16} - 10^{14}$ cm$^{-3}$, 
the inversion layer goes up to $1500 - 15000$ {\AA}, precluding the observation of ballistic holes. 
In conclusion, high doping and a thin metal base are critical factors in detecting holes, as demonstrated by our experiments. 

As described in Section~\ref{sct:thPH}, for Au/Ge and ballistic holes we use $\mu=2.4$. We include the effect of the electron-phonon interaction 
by adding $\mu'=4$, as described in Equation~\ref{eqn:3}.
Here we find a single onset at 
$\overline{\Phi}=-0.58\pm 0.05$ eV (Table~\ref{tbl:VnIn}). 
However, caution is in order; intensities for hole injection in the valence band are about $\frac{1}{100}$ weaker than for injection of ballistic electrons in the conduction band. The noise-to-signal ratio is correspondingly higher, making it challenging to identify a double threshold that may exist but is not seen. 
Since the Fermi level is below the conduction band minimum by $0.01$ to $0.035$ eV for 
some doping between $10^{18}$ to $10^{16}$ cm$^{-3}$, we expect the direct barrier for
electrons to exceed the reverse barrier for holes by that amount. 
Of the two barriers found for electrons, the second one dominates
for voltages above that second threshold. 
Therefore, we tentatively associate the single barrier we have identified for holes with the second barrier for electrons, giving a minimum difference of $\Phi_{e}-\Phi_{h}=0.035$ eV, which is inside the error bars.

\begin{table}[!t]
\begin{tabular}{|c|c|c|} \hline
S            &$\Phi_1$           &    $R^2$          \\ \hline       
1            &  $-0.58\pm 0.04$  &    0.86 \\     
2            &  $-0.58\pm 0.05$  &    0.9895 \\  
A            &  $-0.58\pm 0.05$  &           \\ \hline
\end{tabular}
\caption{Reverse polarization injection of holes into the 
semiconductor valence band ($V<0$ and $I<0$) for spectra S 1 and 2.
Symbols as in Table~\ref{tbl:VpIp}.
The error for the average is taken as the largest individual standard error
in the fit.}
\label{tbl:VnIn}
\end{table}

\subsubsection{Injection of Auger-like electrons into the semiconductor conduction band}
\label{sct:Vm0I0}


\begin{table}[!b]
\begin{tabular}{|c|c|c|c|c|} \hline
S   &$\Phi_1 (eV)$  & $\Phi_2$  (eV)   &  n  & $R^2$ \\ \hline                
3   & $0.550\pm 0.001$           &  $0.670\pm 0.001$              & $10^{18}$  & 0.99999\\ 
4   & $0.550\pm 0.002$           &  $0.635\pm 0.003$              & $10^{16}$ & 0.99999\\ 
5   & $0.599\pm 0.002$           &  $0.720\pm 0.001$              & $10^{14}$ &  0.99998\\  
A   & $0.57\pm 0.03$   & $0.68\pm 0.04$ &  & \\ \hline
\end{tabular}
\caption{Reverse polarization injection of Auger-like electrons into
the semiconductor conduction band  ($V<0$ and $I>0$) for spectra S 3, 4 and 5;
different dopings $n$ are given in cm$^{-3}$.
Other symbols as in Table~\ref{tbl:VpIp}.}
\label{tbl:VnIp}
\end{table}

Finally, the width of the metal base determines not only the attenuation in that region but also the rate of creation for secondary Auger-like electrons near the onset, as can be seen in Figure~\ref{fgr:fig1}-b comparing the blue and red curves where
$I_b$ changes from negative to positive for $V \le \Phi <0$. Doping of the semiconductor and the corresponding marginal layer's width also plays a role since the intensity due to holes,  in the opposite direction to Auger electrons, depends on these parameters. 
However, comparing results for samples 3 to 5 in 
Figures~\ref{fgr:M08S3},~\ref{fgr:M02S4}, and~\ref{fgr:M28S5}, we conclude that
the main parameter controlling the generation of Auger electrons corresponds to the width of the metal base. 
Given some mean free path $\lambda$, the probability for having an Auger process in some distance $x$ can be estimated as $P(x)=1-e^{-\frac{x}{\lambda}}$. The ratio of probabilities between a wide metal base, e.g. $m~x$, to a thin one, $x$, can 
be expanded for small x to obtain $R=\frac{P(mx)}{P(x)} \approx m + \frac{m}{2} (1-m) x$. To form an estimate we take $\lambda = 400$ {\AA} (mean free path
for inelastic processes in Au) and $x=50$ {\AA}, which suggest that a wider metal base $m~x$ can have about $m$ times more probability to generate Auger processes, and therefore a corresponding
stronger secondary current of electrons.
Such is the case we observe in spectra 3, 4 and 5.

Since the relevant phase space here is related to the conduction band, we use again the value 
$\mu=2.12$ for injection of ballistic electrons 
and we add the probability of the inelastic loss that could create an Auger-like secondary electron, $\mu'=2$.
Such value is close enough to the power $n=4$ we have used in Figure~\ref{fgr:fig1} to analyse the probability distribution of two barriers using the Bell-Kaiser model ($T= 0$ K), and it validates {\it a posteriori} the use of such an approximation.
We recover two onsets, 
$\overline{\Phi_{1}}=0.57\pm 0.03$ and $\overline{\Phi_{2}}=0.68\pm 0.04$ eV,
which are compatible with our previous findings for the injection of elastic ballistic electrons. 
The variation in the onsets with the different dopings is inside error bars, and we cannot confidently conclude about the dependence of the Schottky barrier with the semiconductor doping.
However, the tendency for the onset values is compatible with what is expected. As revealed by k-space ab-initio Monte-Carlo simulation, electron-electron inelastic interaction opposes the focusing effect of diffraction due to the propagation through a periodic lattice and tends to defocus beams of 
secondary Auger electrons\cite{hohenester01,DEANDRES20013}.  
However, propagation of these
electrons by more than ten layers of the metal base restore 
wavefunctions similar to the ones for ballistic electrons
via the elastic interaction with the metallic periodic lattice,
making these secondary electrons quite similar to the primary ones.

Finally, the semiconductor can also sustain Auger processes, but
only for higher voltages, $V \ge \Phi+E_G$ which are not relevant in our
best-fit determination of $\Phi$ that only uses a small interval of voltages
near the onset. Therefore, these processes are not considered here.

\section{Conclusions}

We have developed an ab-initio phase-space formalism to describe currents of holes injected at negative voltages and secondary electrons
formed via inelastic Auger processes. To that end, we have included the role of phonons interacting with holes probing the valence band and the Auger-like secondary beam, probing the conduction. 
Except for the ab-initio determination of the effective phase-space volume around the conduction band minima or the valence band maxima, the formalism is simple enough, as shown by Equation~\ref{eqn:3}.
It allows an accurate parameter-free determination of the onsets
that characterize the Schottky barrier. Such an advantage over phenomenological models allows extracting Schottky barrier information from BEES in a technologically interesting but not-trivial situation, namely the two barriers found for injection of electrons in the conduction band in Au/Ge.

For the particular Au/Ge interface, we have found
that direct injection of electrons into the conduction band shows two onsets 
at $\overline{\Phi_{1}}=0.59\pm 0.03$ 
and $\overline{\Phi_{2}}=0.72\pm 0.02$ eV, which agrees with our previous findings. 
Secondary Auger-like electrons injected into the conduction band
also show two onsets at
$\overline{\Phi_{1}}=0.59\pm 0.01$ 
and $\overline{\Phi_{2}}=0.70\pm 0.02$ eV, also 
in good agreement with our previous findings. 
However, for the injection of holes into the valence band, we have
identified only a single onset at 
$\overline{\Phi}=-0.58\pm 0.01$ eV. Given the weakness of those currents, we cannot confidently conclude that this is a physical effect rather than the consequence of the higher noise-to-signal ratio.

{\bf Acknowledgments.} This work was supported by 
the Spanish Ministry of Science (PID2020-113142RB-C21, MCIN/AEI/10.13039/501100011033).
PdA acknowledges a Mobility Grant from the Spanish Ministry of Education. 

\bibliographystyle{apsrev}
\bibliography{beem}

\newpage
\appendix

\section{Supplementary Information}

\subsection{Ab-initio determination of $\mu$ for hole injection}
\label{sct:muh}

\begin{figure}[!h]
\centering
\includegraphics[width=0.99\linewidth]{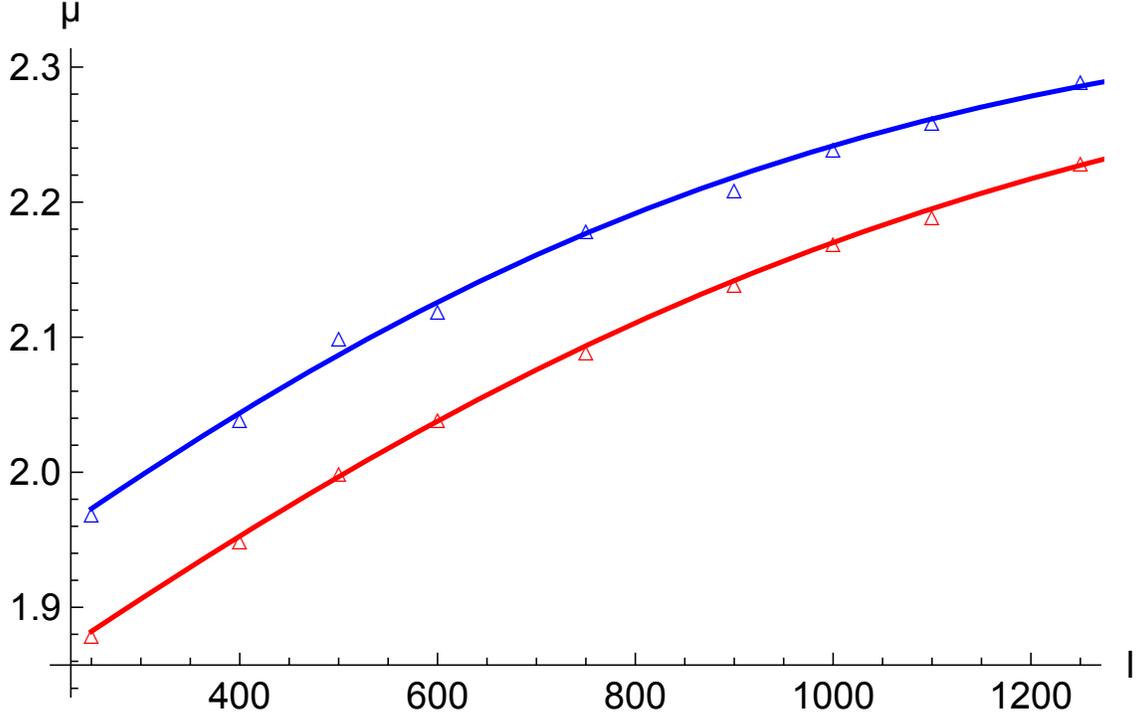}
\caption{$\mu$ derived from first-principles calculations for 
injection of holes as a function of the number of tetra-layers ($l$)
used in the simulation to build up the semiconductor. 
Extrapolated values using the continuous fitting lines for
an infinite number of layers
give $\mu=2.36\pm 0.05$, where the error bar is
derived by comparing the values obtained for two 
voltage intervals ($\Phi+0.25$ and $\Phi+0.35$ eV; blue and red, 
respectively).}
\label{fgr:muh}
\end{figure}

To determine the effective phase-space volume related to the injection of holes in the semiconductor valence band, we use our ab-initio approach to compute BEES I(V) for the interface Au/Ge under reverse bias conditions~\cite{gerbi2018}. A critical point in this calculation is the band bending that appears on the semiconducting side. To simulate this effect, we have included a large number of layers ($>1000$ {\AA}) in the description of the semiconductor. The large depletion layer is connected to a bulk-like structure, obtained with the decimation technique~\cite{claveau2017}, and the bending of bands has been modelled using quadratic interpolation. In our approach, the electronic states of each layer were shifted layer by layer in small energetic steps from the original position of the Fermi level at the interface to the final energy
at the end of the intermediate area just before the semiconducting bulk. The resulting BEES curve was then fitted in the same way as previously explained for the injection of electrons~\cite{gerbi2018}. For the pure ballistic current of holes (virtually no attenuation and $T=0$ K), we obtain for Au/Ge $\mu=2.4$, cf. Figure~\ref{fgr:muh}.

\subsection{Modification of $\mu$ due to phonons}
\label{sct:muph}

Next, we study the attenuation of the ballistic current due to phonons which are the main source of attenuation in the low-energy region, as commented in Section~\ref{sct:thPH}.

Let us consider the propagation through the marginal layer of a hole injected at the top of the valence band, $E=0$.
Any loss of energy due to quasi-elastic interactions with
phonons should prevent the hole from getting to the bulk of the semiconductor.
The probability for such a hole to traverse the width of the marginal layer, $L$, without interacting with phonons is $e^{-p}$,
where $p=\frac{L}{\lambda}$ is the average rate of scattering with phonons determined by the mean free path $\lambda$. Therefore,
we apply such probability as an attenuation factor for holes with energy
in the interval $-w<E<0$, with $w$ a typical frequency for optical phonons ($w=0.035$ eV).

We are interested in a small interval of energies near the onset 
which is helpful to obtain a best fit for the value for the Schottky barrier. 
In such an interval, e.g. $(w,5w)=(0.01,0.18)$ eV, holes interacting with more than five phonons do not have enough energy to be injected into the valence band and can be neglected. 
Furthermore, in such a small interval of energies
$\lambda(E)$ takes values that vary slowly and
for the sake of simplicity can be approximated by its averaged
value over that interval, 
$\overline{\lambda}=\frac{1}{4w}\int_{w}^{5w} \lambda(E)~d~E= 15$ {\AA}.

\begin{figure}[!h]
\centering
\includegraphics[width=0.99\linewidth]{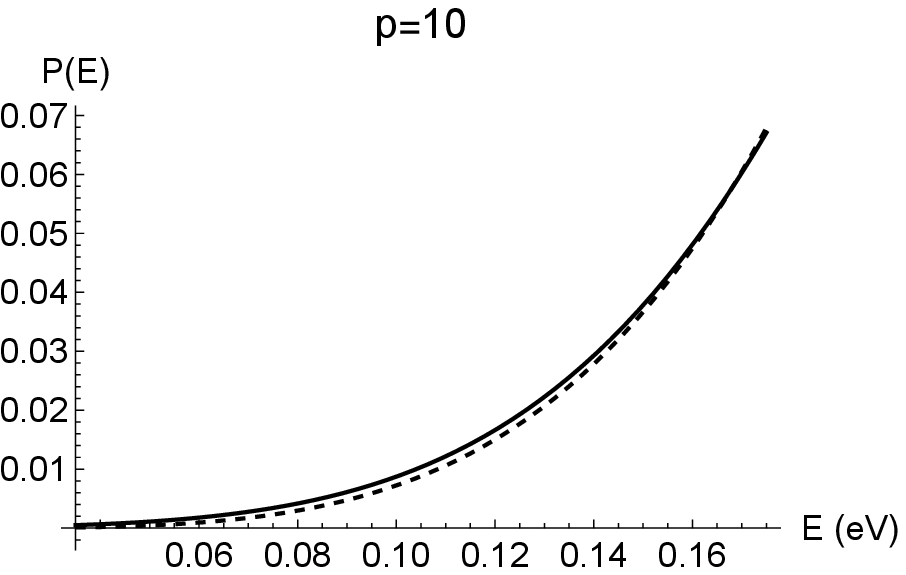}
\caption{Black continous line: $P(p,N)$ for $p=10$ and $N=\floor*{E=\frac{V-\Phi}{w}}\approx 5$
in the interval $(w,5w)$ with $w=0.035$ eV. The dashed line corresponds to
the interpolating power-law approximation $72.4~E^4$.}
\label{fgr:muppp}
\end{figure}

We model the probability of the carrier interacting with $n$ phonons 
while being propagated through the marginal layer
by a Poisson distribution, 
\begin{equation}
P(p,N)= e^{-p}~\sum_{n=0}^{N}\frac{(p)^{n}}{n!}~ ,
\end{equation}
\noindent
where $p=\frac{L}{\lambda}\approx 10$ is the average mean rate
of phonons expected in
the length $L$ and 
$N=\floor*{E=\frac{V-\Phi}{w}}\approx 5$ is the greatest integer less than or equal to $\frac{V-\Phi}{w}$, i.e. the maximum number of possible phonons for each $V$.
Figure~\ref{fgr:muppp} compares $P(p,N)$ with the power-law
interpolating function $72.4~E^4$, which justifies
the use of $\mu'=4$ up to $V<0.18$ eV 
to describe the quasi-elastic interaction between holes and phonons.

\subsection{Modification of $\mu$ due to Auger electrons}
\label{sct:tauQP}

\begin{figure}[!h]
\centering
\includegraphics[width=0.99\linewidth]{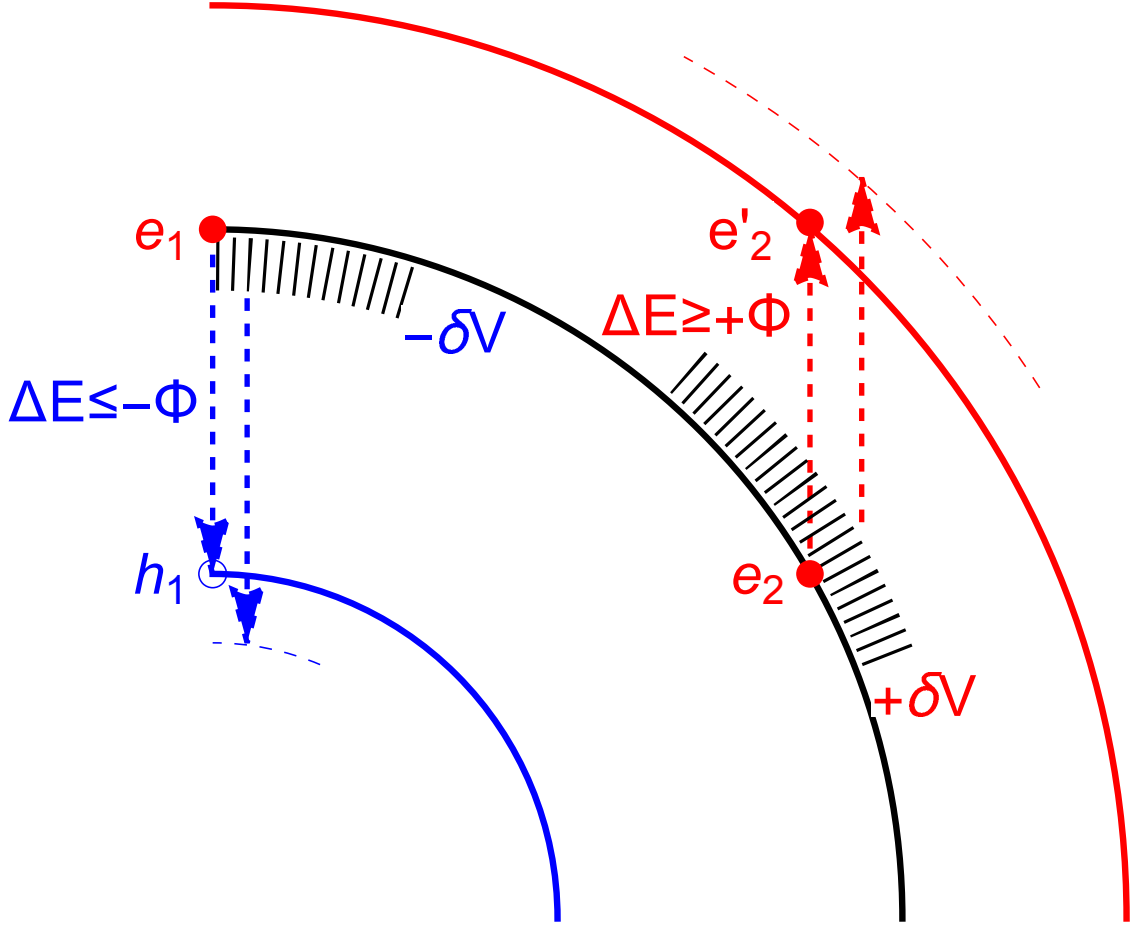}
\caption{k-space scheme for the generation of
secondary Auger electrons in the metal base
as explained in detail in~\ref{sct:tauQP}. Notice that
the shadowed area is proportional to $\delta V$.}
\label{fgr:muAugerM}
\end{figure}

Figure~\ref{fgr:bands} (bottom-left) shows an energy scheme
for the generation of Auger electrons in the metal. 
A ballistic hole $h_1$ travelling with enough energy to
be injected near the top of the valence band of the semiconductor
($V\le \Phi$) interacts inelastically with an electron $e_1$ near the Fermi surface transfering that energy to another electron $e_2$, also near the Fermi surface, which uses the transferred energy
to form a secondary electron $e'_{2}$
with enough energy to be injected near the minimum of the
conduction band. 
A qualitative argument based on the scheme displayed in
Figure~\ref{fgr:muAugerM} shows how simultaneous conservation
of momentum and energy for
electrons $e_1$ inside an spherical shell of width
$E_F<E<E_F-\delta V$ recombining with holes $h_1$,
can transfer the right momentum and energy to electrons $e_2$ 
inside a spherical shell of width
$E_F<E<E_F+\delta V$, to get enough energy to be injected
near the minimum of the conduction band. 
Such a process increases the phase-space volume by
the width of some spherical shell of width $\delta V$ for $e_1$ and again for $e_2$,
hence $\mu'=2$. 
This qualitative argument agrees with the probability
of inelastic interaction between electrons $e_1$ and $e_2$ 
so $e_1$ loses energy $\approx \Phi+\delta$ decaying to $h_1$ 
and $e_2$ can use it to be promoted to level $e'_{2}$.
Such process
has been computed long ago using the low-energy excitation spectra of
free-electron metals by Quinn~\cite{Quinn}. A full 
ab-initio calculation for Au and Pd by 
Ladst{\"a}dter et al.~\cite{ladstadter2003} confirms  
Quinn's arguments for free-electrons metals like Au.
It becomes proportional to,

\begin{equation}
\tau^{-1} (\frac{E}{E_F},r_s) \approx \frac{10+r_s^{\frac{5}{2}}}{600}
\frac{\left(E - E_F \right)^2}{\sqrt{E}}~~(\text{fs}^{-1})~;~1<r_s<5
\label{eqn:tau}
\end{equation}
\noindent
This value can be interpreted as the rate of energy loss due to Coulomb interaction,
and yields again $\mu'\approx 2$ for $V=E-E_F<E$
in agreement with previous qualitative arguments in the 
literature~\cite{bellJVST91,ludeke93JVST}.

For the sake of completitude, 
Figure~\ref{fgr:bands}-(bottom-right) shows the corresponding 
energy scheme
for the generation of Auger electrons in the semiconductor. 
Since this case only contributes to the current
for $V\ge \Phi+E_G$, which is outside the interval of
interest for our best-fit determination of $\Phi$,
we do not include it in our calculations.

\subsection{Ge}
\label{sct:nGe}

Here, we compute values for Ge at $T=80$ K and different doping leves that have
been used in the paper.~\cite{sze}

For intrinsic Ge, the number of electrons in the conduction band at a given
T is,

\begin{equation}
n=\int_{E_C}^{\infty} N(E) F(E) dE = N_C \frac{2}{\sqrt{\pi}} F_{\frac{1}{2}} (x_F)
\label{eqn:nC}
\end{equation}
\noindent
where $F_{\frac{1}{2}} (x_F)=\int_{0}^{\infty} 
\frac{ x^{\frac{1}{2}}~d~x}{1+e^{(x-x_F)}}$ is the Fermi-Dirac integral,
$x_F=\frac{E_F-E_C}{k T}<0$ and,
$N_C$ is the effective density of states in the conduction band (T= 80 K),
\begin{equation}
N_C= 2 M_C \left(\frac{2\pi m_{de} k_B T}{h^2} \right)^{\frac{3}{2}}
= 1.452\times 10^{18} \text{cm}^{-3} 
\label{eqn:NCT}
\end{equation}
\noindent
where $M_C$ is the number of equivalent minima in the conduction band
($M_C=4$ for Ge) and $m_{de} = (m_1 m_2 m_3)^{\frac{1}{3}}$ is the
density of states effective mass for electrons
($m_{de}=0.223 m_0$ for Ge, with $m_0$ the electron mass in vacuum).
The number of holes near the top of the valence band, $p$, is given
by similar expressions, with 
$N_V= 8.158\times 10^{17} \text{cm}^{-3}$ (T= 80 K).

The intrinsic Fermi level is obtained by using $n=p$. We have (T= 80 K),

\begin{equation}
E_{F_i}=\frac{E_C+E_V}{2} + \frac{3~k_B T}{4} \ln{\frac{N_V}{N_C}} = 0.364 \text{eV}
\label{eqn:EFi}
\end{equation}
\noindent
where midgap is $\frac{E_C+E_V}{2} = 0.37185 ~\text{eV}$.
The variation of the gap with temperature is described by
the simple function, 
$E_G=0.7437 - 4.774\times 10^{-4} T^2/(T + 235)$,
which gives $0.734$ for $T=80$ K.

Finally, the intrinsic carrier density is
$n_{i}=\sqrt{n \times p} = \sqrt{N_C \times N_V} e^{-\frac{E_G}{2 k_B T}}
=8.26\times 10^{-6} \text{cm}^{-3}$ (T=80 K). 

For our heavily n-doped samples, $N_D$ is in the range
$10^{18} \text{cm}^{-3}$ to $10^{19} \text{cm}^{-3}$
(0.001-0.01 $\Omega$ cm).
For these values the respective saturation temperatures are 
$T_S=108$ and $318$ K. Therefore, we are operating in the
ionization regime and impurities dominate the conductivity of
the semiconductor. 

In the doped material the Fermi level adjusts to preserve
charge neutrality, $n = p + N_D^{+}$.
Assuming that donor impurities with
a concentration $N_D$ are located at $\Delta E=E_D$ below the
conduction band ($E_D=0.0096$ eV for Au/Ge) we have,

\begin{equation}
N_D^{+} 
=\frac{N_D}{1+2~e^{\frac{E_F-E_D}{k_B T}}}
\end{equation}
\noindent

Therefore, the charge neutrality condition reads,
\begin{equation}
N_C e^{-\frac{E_C-E_F}{k_B T}} =
N_V e^{+\frac{E_V-E_F}{k_B T}} +
\frac{N_D}{1+2~e^{\frac{E_F-E_D}{k_B T}}}
\label{eqn:EFEX}
\end{equation}
\noindent
which can be solved numerically, 
to obtain:
$E_{F_n}=0.724$ eV for $N_D=10^{18}~\text{cm}^{-3}$,
$E_{F_n}=0.6995$ eV for $N_D=10^{16}~\text{cm}^{-3}$, and
$E_{F_n}=0.668$ eV for $N_D=10^{14}~\text{cm}^{-3}$.


Finally, we use Debye's length,
$L_D = \sqrt{\frac{\epsilon_S k_B T}{q_e^2 N_D}}$ to estimate the
size of the corresponding depletion layer.
For Ge, $\epsilon_S = 16.0~\epsilon_0$ 
($\epsilon_0$ the vacuum permitivity).
At $T=80$ K we get $L_D=25$ {\AA} for $N_D=10^{18}$
and $L_D=8$ {\AA} for $N_D=10^{19}$. 
For an abrupt junction, the inversion layer in Ge is
estimated to be about $L_I \approx 6~L_D$.

The Fermi level moves from the low to the high
doping sample closer to the conduction band by $0.056$ eV.
This effect has not been seen in the corresponding values 
for the Schottky barrier because it falls inside error bars.

\subsubsection{Mean free path of holes in Ge}
\label{sct:lamGe}
\begin{figure}[!h]
\includegraphics[width=0.99\linewidth]{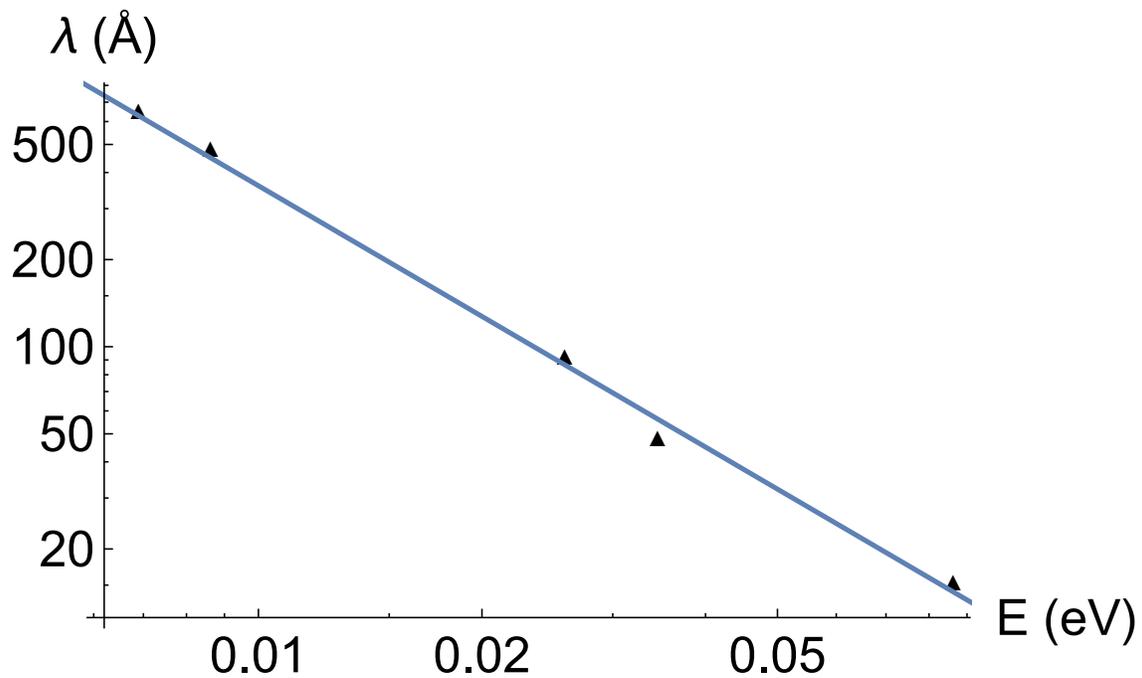}
\caption{Mean free path as a function of the energy of carriers, $\lambda(E)$ ({\AA}),
for holes injected near the top of the valence band ($E=0$) in Ge.
Triangles are derived from mobility values reported by Brown and Bray~\cite{brown1962}.
The continous line is an interpolation function, $\lambda(E)=\frac{0.36}{E^{1.5}}$,
given to help to guide the eye.
}
\label{fgr:lamGe}
\end{figure}

We derive values for the low-energy mean free path of holes injected in the semiconductor valence 
using values for the mobility $\mu$ reported by Brown and Bray~\cite{brown1962},
\begin{equation}
\lambda = \frac{m}{e} v \mu
\end{equation}
\noindent
Where $m$ and $e$ are the mass and charge of holes in the valence band and
$v=\sqrt{\frac{2E}{m}}$ is a typical velocity used to convert lifetimes in mean free paths.
Figure~\ref{fgr:lamGe} shows how the mean free path changes from 
about $55$ {\AA} for the typical optical phonon frequency of $w=0.035$ eV
to about $5$ {\AA} for the energy of five phonons, $0.18$ eV,
which determines the interval of energies used in our analysis to 
determine the Schottky onset.

\subsection{Best fits}

\begin{figure}[!h]
\includegraphics[width=0.99\linewidth]{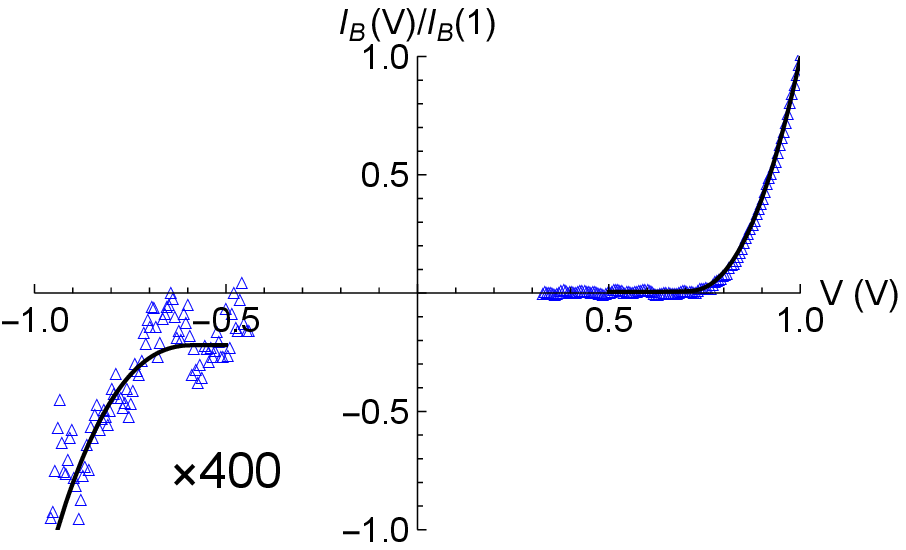}
\includegraphics[width=0.49\linewidth]{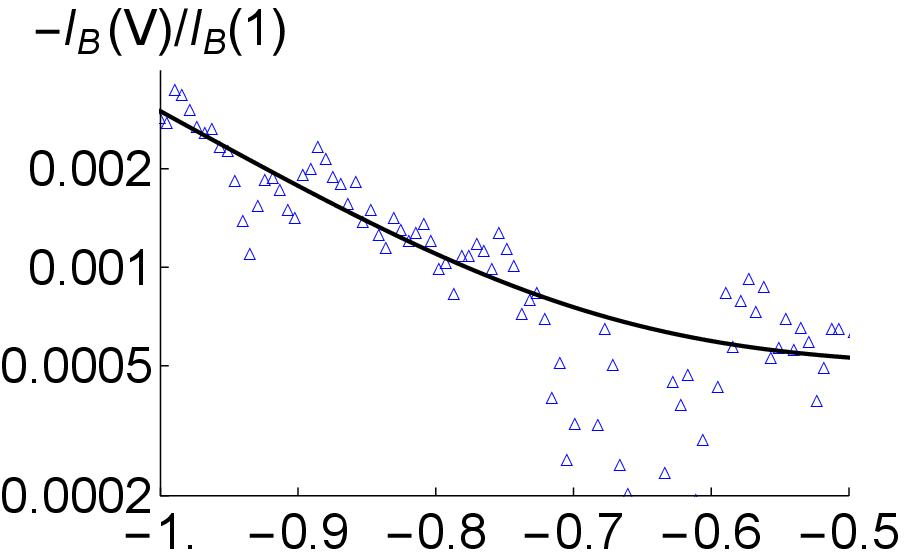}
\includegraphics[width=0.49\linewidth]{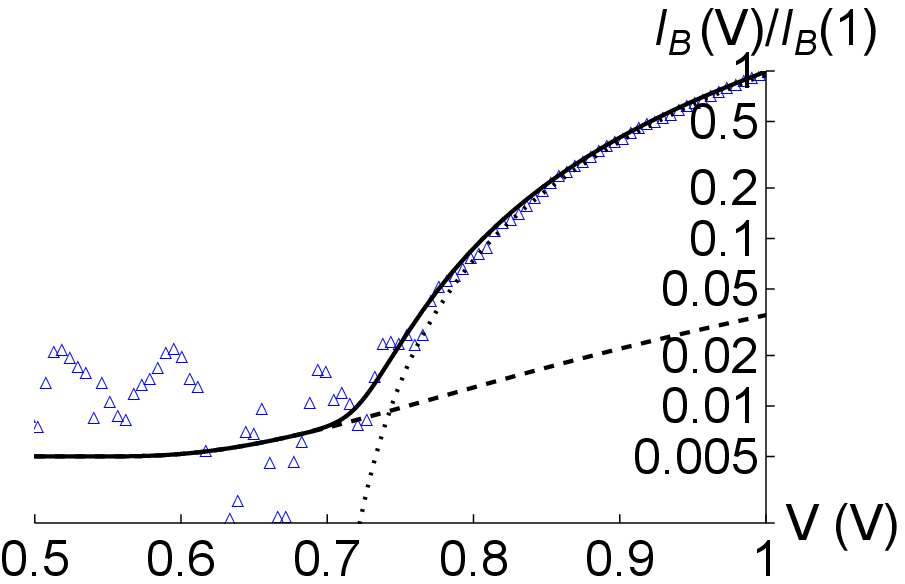} 
\caption{Upper panel:
Best fit using Equation~\ref{eqn:3} (black continous line)
to experimental BEES I(V)'s (blue triangles) for spectrum 1 ($I_t=2$ nA).
The reverse current (weak injection of holes for $V<0$)
has been multiplied by a factor $400$ to show it
on the same scale as the direct current (injection of electrons for $V>0$).
Lower panels: I(V) is shown on a logarithmic scale to highlight the
behaviour near the onset. For the cases where two onsets have been
identified, dashed and dotted lines show the
two components used to get the best fit to data.
}
\label{fgr:M36S1}
\end{figure}

\begin{figure}[!h]
\includegraphics[width=0.99\linewidth]{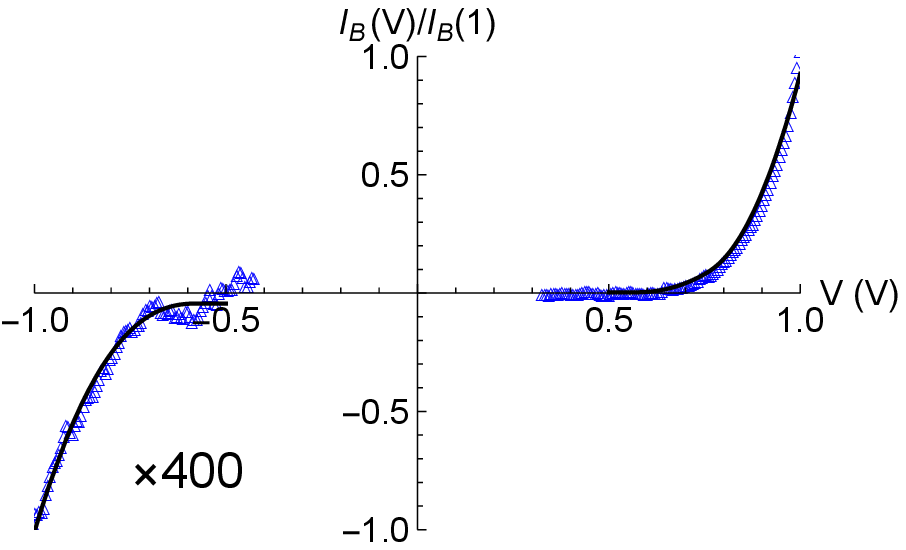}
\includegraphics[width=0.49\linewidth]{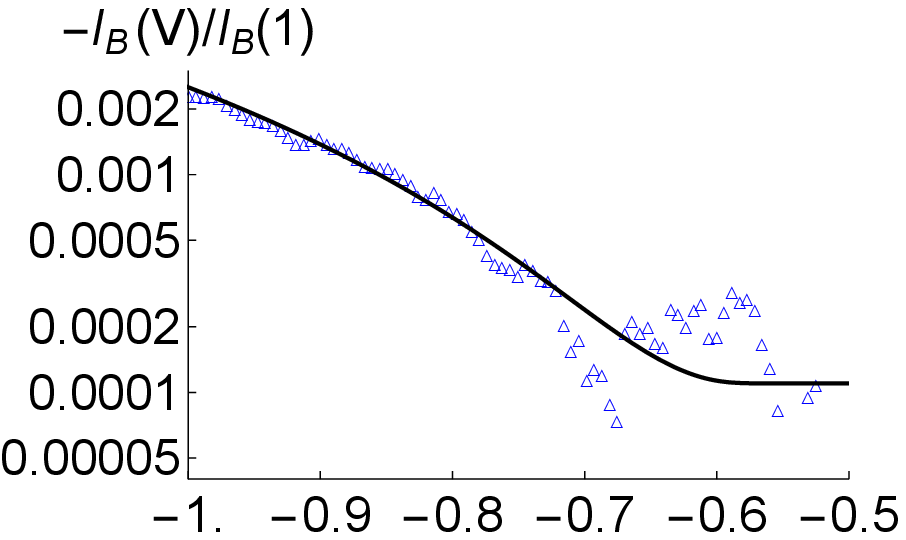}
\includegraphics[width=0.49\linewidth]{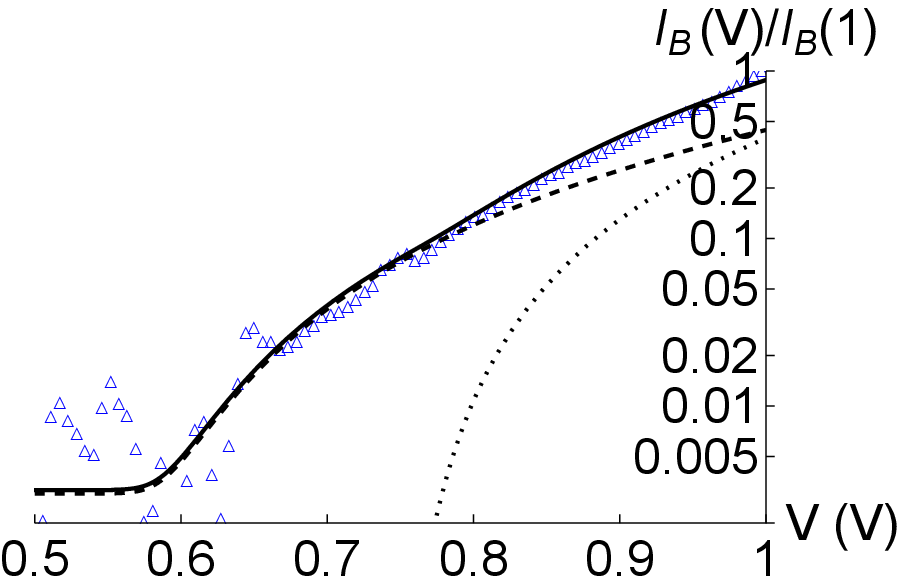} 
\caption{As in Figure~\ref{fgr:M36S1}, for spectrum 2 ($I_t=10$ nA).
}
\label{fgr:M42S2}
\end{figure}

\begin{figure}[!h]
\includegraphics[width=0.99\linewidth]{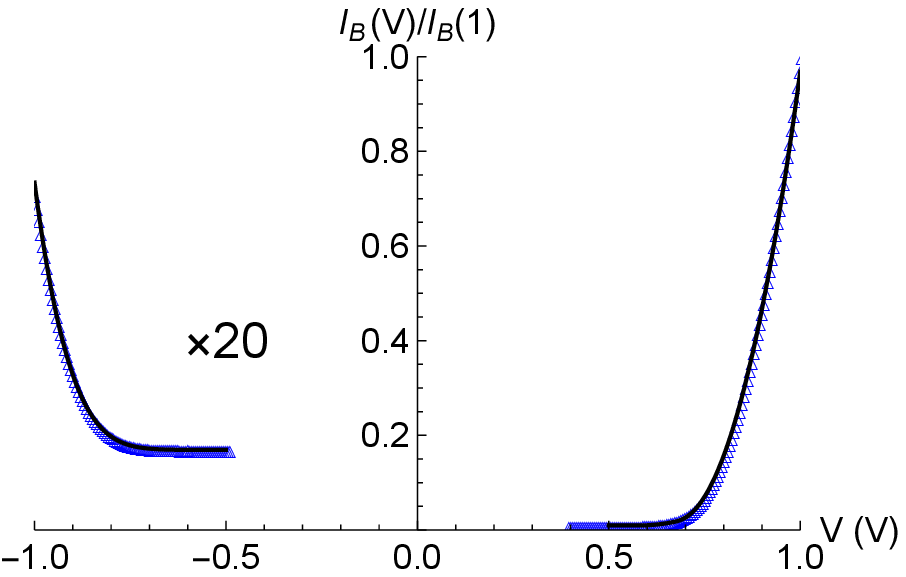}
\includegraphics[width=0.99\linewidth]{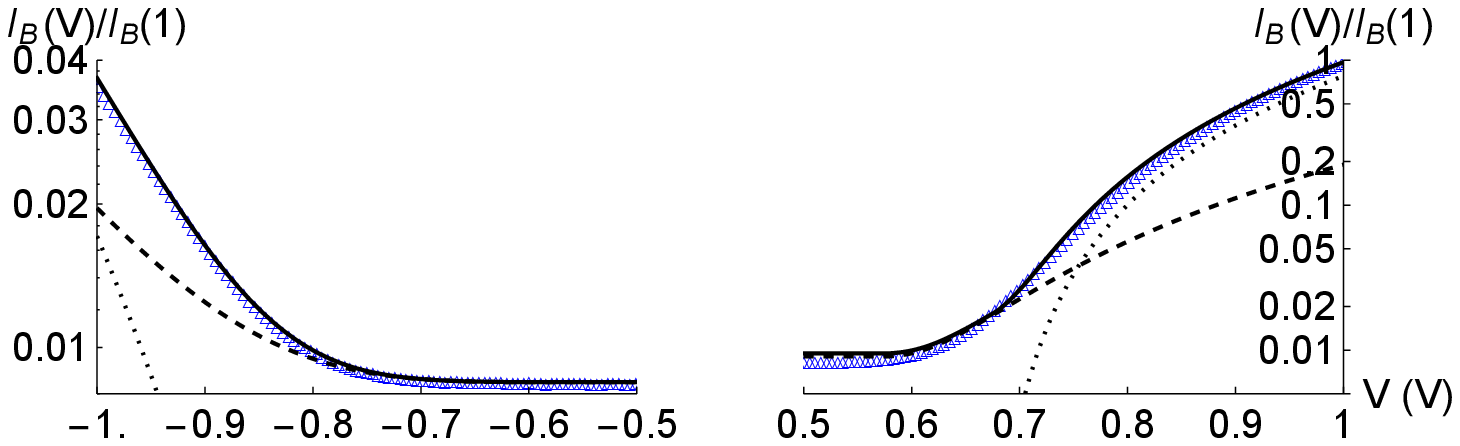}
\caption{As in Figure~\ref{fgr:M36S1}, for spectrum 3 ($I_t=20$ nA).
The scale factor used to show positive and
negative voltages on a similar scale is now $20$.
}
\label{fgr:M08S3}
\end{figure}

\begin{figure}[!h]
\includegraphics[width=0.99\linewidth]{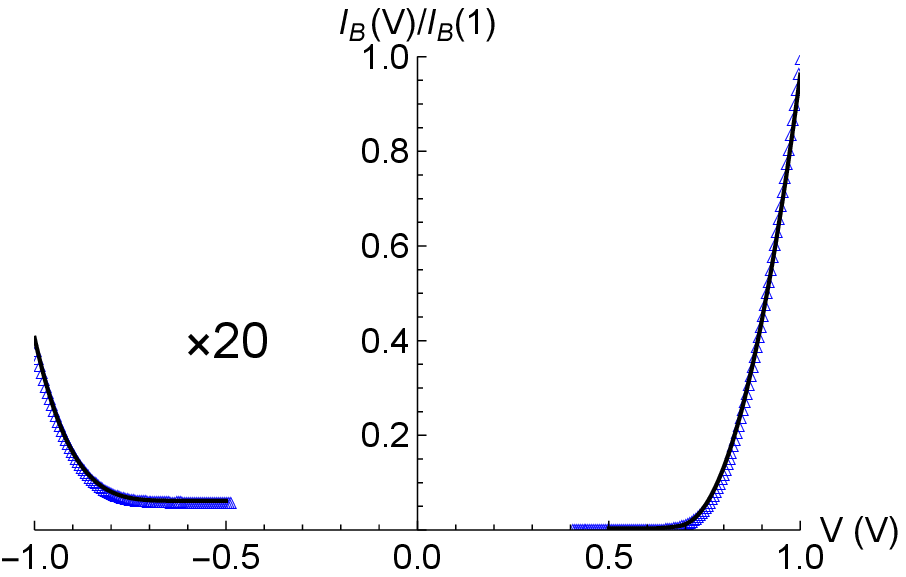} 
\includegraphics[width=0.99\linewidth]{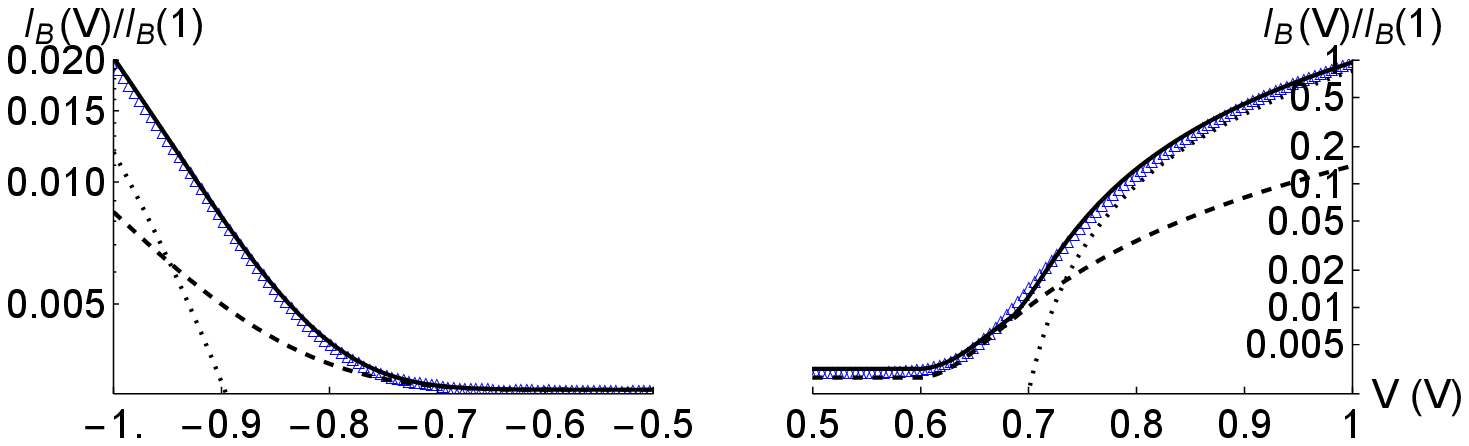}
\caption{As in Figure~\ref{fgr:M08S3}, for spectrum 4 ($I_t=20$ nA).
}
\label{fgr:M02S4}
\end{figure}

\begin{figure}[!h]
\includegraphics[width=0.99\linewidth]{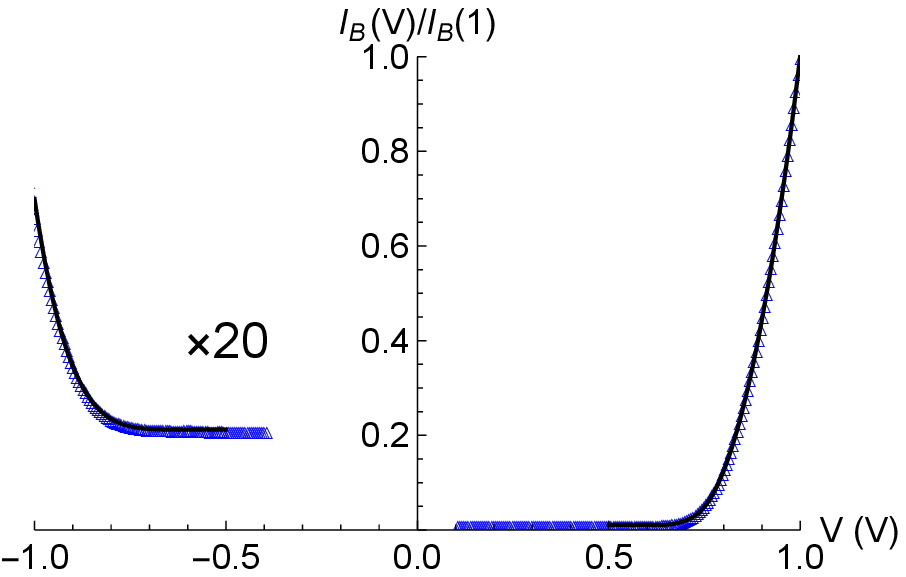} 
\includegraphics[width=0.99\linewidth]{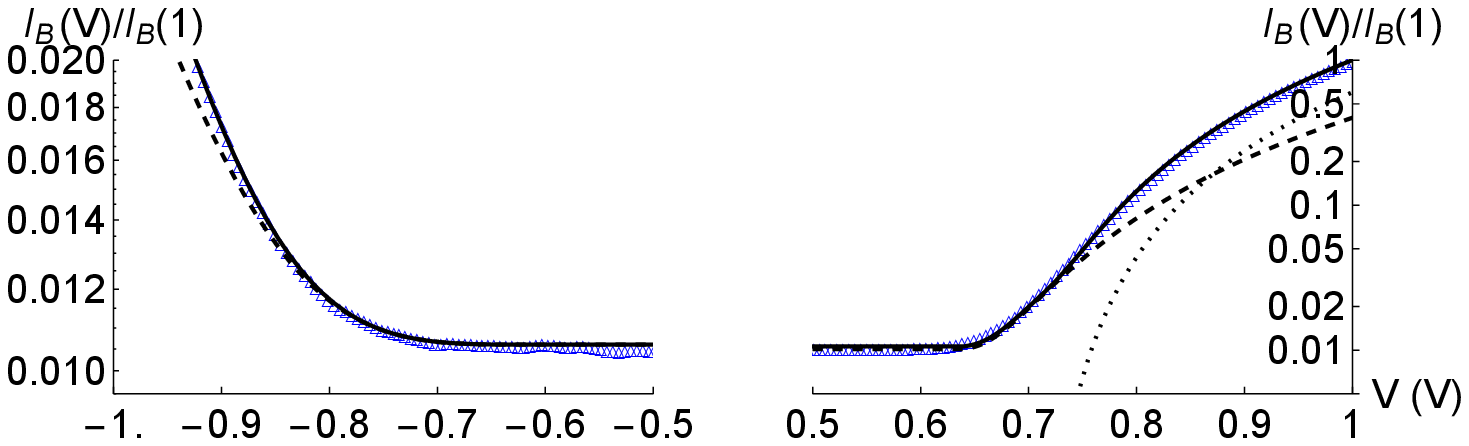}
\caption{As in Figure~\ref{fgr:M08S3}, for spectrum 5 ($I_t=2$ nA).
}
\label{fgr:M28S5}
\end{figure}

Best fits for spectra 1 to 5 are shown in
Figs.~\ref{fgr:M36S1},~\ref{fgr:M42S2},~\ref{fgr:M08S3},~\ref{fgr:M02S4} and,~\ref{fgr:M28S5}.
The central sub-threshold region has been modelled as a constant noisy
contribution that has been calculated by a simple average over the region,
taken separately for positive and negative voltage.
To facilitate the analysis 
the sub-threshold offset has been
subtracted from the experimental data,
and intensities have been normalized to the value $I_B (V=1)$
to bring all data on a similar scale.

\end{document}